\documentclass[aps,prl,showpacs,amsmath,amssymb,amsfonts,lengthcheck,superscriptaddress,twocolumn,longbibliography]{revtex4-1}

\usepackage[dvipsnames]{xcolor}
 
\usepackage{amssymb}
\usepackage{color}

\usepackage{graphicx}
\usepackage{subfigure}
\usepackage{amsthm}
\usepackage{verbatim}
\usepackage{dcolumn}
\usepackage{bm}
\usepackage{epsf}
\usepackage[colorlinks=true,citecolor=blue,pdfusetitle,linkcolor=blue,urlcolor=blue]{hyperref}%
\usepackage{dsfont}

\makeatletter

\@ifundefined{textcolor}{}
{%
 \definecolor{BLACK}{gray}{0}
 \definecolor{WHITE}{gray}{1}
 \definecolor{RED}{rgb}{1,0,0}
 \definecolor{GREEN}{rgb}{0,1,0}
 \definecolor{BLUE}{rgb}{0,0,1}
 \definecolor{CYAN}{cmyk}{1,0,0,0}
 \definecolor{MAGENTA}{cmyk}{0,1,0,0}
 \definecolor{YELLOW}{cmyk}{0,0,1,0}
}

\newcommand{\blah}{blah\\blah\\blah\\blah\\blah.}

\newcommand{\ket}[1]{\left\vert#1\right\rangle}
\newcommand{\bra}[1]{\left\langle#1\right\vert}
\newcommand{\braket}[2]{\left\langle #1|#2\right\rangle}

\newcommand{\com}[2]{\left[#1,\,#2\right]}

\newcommand{\bla}{bla\\bla\\bla\\bla\\bla}

\newcommand{\mrm}[1]{\mathrm{#1}}

\begin{document}

\title{Kibble-Zurek scaling in quantum speed limits for shortcuts to adiabaticity}

\author{Ricardo Puebla}
\email{r.puebla@qub.ac.uk}
\affiliation{Centre for Theoretical Atomic, Molecular and Optical Physics, School of Mathematics and Physics, Queen’s University Belfast, Belfast BT7 1NN, United Kingdom}
\author{Sebastian Deffner}
\email{deffner@umbc.edu}
\affiliation{Department of Physics, University of Maryland, Baltimore County, Baltimore, MD, 21250, USA}
\author{Steve Campbell}
\email{steve.campbell@ucd.ie}
\affiliation{School of Physics, University College Dublin, Belfield Dublin 4, Ireland}

\begin{abstract}
Geometric quantum speed limits quantify the trade-off between the rate with which quantum states can change and the resources that are expended during the evolution. Counterdiabatic driving is a unique tool from shortcuts to adiabaticity to speed up quantum dynamics while completely suppressing nonequilibrium excitations. We show that the quantum speed limit for counterdiabatically driven systems undergoing quantum phase transitions fully encodes the Kibble-Zurek mechanism by correctly predicting the transition from adiabatic to impulse regimes. Our findings are demonstrated for three scenarios, namely the transverse field Ising, the Landau-Zener, and the Lipkin-Meshkov-Glick models.
\end{abstract}

\maketitle
A particularly promising approach to quantum computing relies on quantum annealing \cite{Kadowaki1998,Harris2010,Johnson2011,Boixo2013}. In this paradigm,  which has been dubbed  \emph{adiabatic quantum computing} \cite{Albash2018}, a quantum system is initially prepared in the ground state of a well-controlled Hamiltonian $H_{\rm i}$. Then, the ``computer'' is let evolve adiabatically -- infinitely slowly -- towards the ground state of the final Hamiltonian $H_{\rm f}$, which encodes the desired solution to the computation. Like all quantum information processing systems, adiabatic quantum devices are subject to the inevitable noise from the environment \cite{Sanders2017} and require quantum error correction \cite{Roffe2019}. However, in adiabatic quantum computing the situation is even more involved than in, e.g., the gate base approach, since computational errors come in two different flavors~\cite{Young2013}: (i) the `usual' errors that are due to the interaction with the environment~\cite{Gardas2018FT} and control noise \cite{Gardas2019PRA}; and (ii), errors that originate in parasitic excitations of the finite-time driving of any realistic system. While powerful algorithms exist to mitigate the former ~\cite{Pudenz2015,Vinci2018}, the latter are significantly harder to control.

At least from a bird's-eye view, so-called shortcuts to adiabaticity~\cite{STAreview} seem exceptionally well-suited to address this issue~\cite{Santos2015SciRep, Takahashi1, Takahashi2}. A shortcut to adiabaticity is a finite-time, controlled process to obtain the final state that would result from infinitely slow, adiabatic driving. In particular, counterdiabatic driving \cite{Demirplak2003,Berry2009} is designed to keep the time-evolving quantum state on the adiabatic manifold at all times. However, with the exception of a few special scenarios \cite{Deffner2014PRX,Jarzynski2017} the necessary control fields to facilitate the shortcut tend to be highly non-local, and thus of only limited practical use in many body systems \cite{delCampoPRL2012,DamskiJStat,CampbellPRL2015,Victor,Deffner2019book}. In addition, current formulations of  shortcuts to adiabaticity are inadequate as a control technique for mitigating computational errors. Often implementing any shortcut in lattice systems at least requires knowledge of the initial and final eigenspectrum, meaning that in order to correct for computational errors the correct outcome has to be known \cite{Deffner2019book}, although recently there has been some efforts to mitigate this requirement through Floquet engineering~\cite{PolkovnikovPRL2019}.

Therefore, alternative strategies and phenomenological approaches, such as linear response theory~\cite{Acconcia2015,deffner2020thermodynamic}, appear instrumental~\cite{PolkovnikovRepPhys}. Within the realm of phenomenology, the Kibble-Zurek mechanism (KZM)~\cite{Zurek2005,delCampo2014} is arguably the most prominent  approach to nonequilibrium quantum dynamics \cite{Dziarmaga2005,DamskiPRL2005,Damski2007,Dziarmaga2012,Francuz2016,Gardas2017,Gardas2018, Adolfo2018PRL, Adolfo2019PRL, Zurek2019PRL, EspositoPRL}. The quantum KZM can be understood as an extension of the quantum adiabatic theorem \cite{born1928adiabatic}. As long as the rate of driving is smaller than the energy gap between ground state and first excited state, barely any transitions occur and the dynamics remains essentially adiabatic. However, close to the critical point of a quantum phase transition (QPT) energy gaps close and excitations become inevitable. The ``amount'' of these excitations can then be predicted entirely from the rate with which the system is driven and from the critical exponents of the QPT~\cite{Zurek2005,Gardas2017}. A rather subtle issue relates to determining precisely when a quantum system transitions between the \emph{adiabatic} and the \emph{impulse} regimes~\cite{DamskiPRL2005,Deffner2017PRE}. As a rule of thumb, this transition is expected to occur when the ``relaxation time'', i.e. one over the energy gap, becomes identical to one over the driving rate~\cite{Zurek2005}. However, the natural question arises whether this expectation can be made more precise.

Recently, it was shown that signatures of the KZM may be present in the quantum speed limit (QSL) \cite{DeffnerReview} when implementing a shortcut to adiabaticity~\cite{CampbellPRL2017}. Thus, while counterdiabatic driving may not be adequate to mitigate all sources of computational errors, the control fields still contain essential information to develop phenomenological quantum error correcting paradigms. Indeed, given that counterdiabatic driving perfectly cancels the excitations that would otherwise occur in the system due to arbitrary ramps, we ask: {\it Can we exploit what is learned from quantum control to quantitatively study the nonequilibrium dynamics we are suppressing?} In the following we answer in the affirmative, showing that quantum control provides a useful window through which nonequilibrium dynamics can be explored. We focus on three systems for which the control fields are analytically known, namely, the transverse field Ising model (TFIM)~\cite{Sachdev,Dutta,Dziarmaga2005,Francuz2016,delCampoPRL2012,DamskiJStat}, the Landau-Zener (LZ) model~\cite{Berry2009} and the Lipkin-Meshkov-Glick (LMG) model~\cite{Lipkin1965,CampbellPRL2015,Campbell2016PRB}, and we examine dynamics approaching and crossing the QPT.

Our results provide a means to quantify the range over which strategies to mitigate fundamentally non-correctable errors are required, i.e. the length of the impulse regime. To this end, we bring together three distinct areas of research, namely shortcuts to adiabaticity, the QSL, and the KZM. We demonstrate that the QSL exhibits a behavior reminiscent of the adiabatic-impulse approximation with distinct minima occurring at the crossover between the regimes. Furthermore, their distance (between left and right of the critical point) is fully consistent with the prediction of the KZM, thus establishing that the speed of the controlled dynamics reveals details of the underlying universality class of the model.

\paragraph{Kibble-Zurek scaling in the quantum speed limit.} 
We start by briefly reviewing the conceptual building blocks and by establishing notions and notations. Consider a time-dependent Hamiltonian $H_0(t)$ with instantaneous eigenvalues $\{\varepsilon_n(t)\}$ and eigenstates $\{\ket{n_t}\}$. An arbitrary evolution of an eigenstate will, in general, lead to non-adiabatic excitations being created. However, we can construct a Hamiltonian, $H(t)=H_0(t)+H_\text{CD}(t)$, such that the adiabatic solution of $H_0(t)$ is the exact solution of the dynamics generated by $H(t)$. Therefore, evolving according to this new Hamiltonian achieves effective adiabatic dynamics in a finite time. The counterdiabatic term (assuming units such that $\hbar\!\!=\!\!1$) is~\cite{STAreview}
\begin{equation}
\label{eq:TQD}
H_\text{CD}(t)= i \com{\partial_t \ket{n_t}\!\!\bra{n_t}}{\ket{n_t}\!\!\bra{n_t}}.
\end{equation}
Implementing a controlled dynamics invariably incurs a thermodynamic toll for suppressing the nonequilibrium excitations and there are several approaches to quantify this energetic cost~\cite{Chen2010PRA, Muga2017PRA, Bravetti2017PRE, Abah2017EPL, MugaNJP2018, Santos2015SciRep, SelsPNAS, LiNJP2018, BarisPRE2019,Abah2019}. A particularly useful measure  proposed by Zheng {\it et al.}~\cite{ZhengPRA2016} relates the instantaneous cost to the norm of the counterdiabatic driving field~\cite{Demirplak2008, ZhengPRA2016}
\begin{equation}
\label{eq:COST}
||H_\text{CD}(t)||\propto \sqrt{\braket{\partial_t n_t}{\partial_t n_t}} \equiv \partial_t C .
\end{equation}
Note that the proportionality constant is dictated by the specific choice of norm employed. For simplicity we neglect this factor and work directly with what is essentially the geometric tensor.

Central to our analysis will be assessing the speed of the evolution. Within the framework of geometric quantum speed limits, a meaningful speed can be defined for any distance measure or metric \cite{DeffnerNJP2017}. For our purposes we shall focus on the norm based approach~\cite{DeffnerPRL2013, DeffnerNJP2017}, for which the QSL for the controlled dynamics is given by~\cite{CampbellPRL2017}
\begin{align}
\label{eq:speednorm}
\nu_{\rm QSL}(t)=\frac{\sqrt{\varepsilon_n^2(t)+(\partial_t C)^2} }{\cos(\mathcal{L}_t)\sin(\mathcal{L}_t)}\,.
\end{align}
Here, $\mathcal{L}_t\!=\!{\rm arccos}|\langle\psi_0|\psi_t \rangle|$ denotes the Bures angle between initial and evolved state at time $t$ and $\partial_t C$ as given in Eq.~\eqref{eq:COST}. Equation~\eqref{eq:speednorm} captures the trade-off between the ``bare" energetic change in the system and the additional resources necessary to achieve the controlled dynamics \cite{CampbellPRL2017}.

At this point, it is natural to question whether our choice of the version of QSL is crucial for the analysis. Over the last decade a plethora of formulations has been proposed~\cite{DeffnerReview}, where QSLs based on the quantum Fisher information~\cite{TaddeiPRL2013,delCampoQSL} give the tightest bound on the actual rate of change. However, such formulations are less useful for studying counterdiabatic driving as the speed along the geodesic of the quantum manifold~\cite{bengtsson2017geometry} is agnostic to the specifics of the nonequilibrium dynamics. In the following, we will focus on the nonequilibrum excitations arising from driving systems through QPTs, and thus for our purposes only formulations of the QSL, such as in Eq.~\eqref{eq:speednorm}, that are sensitive to the full energyspectrum will do. 

Quantum as well as classical phase transitions are characterized by the fact that close to the critical point both the correlation length, $\xi$, as well as the correlation time, $\tau$, diverge. Renormalization group theory predicts \cite{Fisher1974,Herbut2007a,Sachdev}
\begin{equation}
\label{eq02}
\xi(g)=\xi_0 |g-g_c|^{-\nu} \quad {\rm and} \quad \tau(g)=\tau_0|g-g_c|^{-z\nu}\,,
\end{equation}
where $g$ is a dimensionless parameter measuring the distance from the critical point $g_c$, $\nu$ is the spatial and $z$ the dynamical critical exponent. Typically, in thermodynamic phase transitions $g$ denotes the relative temperature \cite{Zurek1985}, whereas in QPTs $g$ is a relative external field \cite{Zurek2005,Francuz2016}. As noted above, for slow-enough driving and far from the critical point, $\tau\!\!\ll \!\! t$, the dynamics of the system is essentially adiabatic. This means, in particular, that all nonequilibrium excitations and defects equilibrate much faster than they are created. Close to the critical point, $\tau\!\simeq \! t$ the situation dramatically changes, since the response ``freezes out'' and defects and excitations cannot ``heal'' any longer.  If the external driving is linear $g(t)\!\!=\!\!t/\tau_q$, and in the conventional phrasing of the KZM \cite{Zurek2005,Zurek1996}, the transition from adiabatic to impulse regime is expected to happen when the rate of driving becomes equal to the rate of relaxation, or more formally at
\begin{equation}\label{eq:AI}
\hat{\tau}(\hat{t})=\hat{t}\quad\mrm{with}\quad\hat{\tau}=\left(\tau_0\,\tau_q^{z \nu}\right)^{1/(z\nu+1)}\,.
\end{equation}
However, this rather hand-waving argument for where to situate the crossover cannot be considered entirely satisfactory. In particular, in unitary quantum dynamics, in which no direct ``relaxation'' can occur, a more rigorous treatment appears desirable.

We therefore propose that the size of the impulse regime is determined by the turning point of Eq.~\eqref{eq:speednorm}, and we identify $\hat t$ from the time when the speed is minimized, $t_m$, cf. Fig.~\ref{figTFIM}(a). In critical systems with time dependent order parameter, $g(t)$, and where Eq.~\eqref{eq:TQD} is known, we find $\partial_t C \!\sim\! \vert\dot{\varepsilon} / \varepsilon \vert$ with $\varepsilon\!\sim\!\vert g - g_c \vert^{z\nu}$ and $g_c$ is the critical point \cite{SM}. In the case of a linear ramp, $g(t)\!\!=\!\!t/\tau_q$ we immediately obtain
\begin{equation}
\label{eq:infi_scaling}
\nu_{{\rm QSL},\text{min}} \sim\tau_q^{-z\nu/(1+z\nu)} \quad {\rm and}\quad  t_m \sim \tau_q^{z\nu/(z\nu+1)} 
\end{equation}
which is precisely the scaling expected from the KZM. In the following we demonstrate this universal result, Eq.~\eqref{eq:infi_scaling}, for two paradigmatic many body systems at the opposite sides of the interaction spectrum -- the transverse field Ising model and the LMG-model.

\paragraph{One dimensional transverse-field Ising model.} We begin with the spin-$1/2$ nearest neighbor Ising model in a transverse field~\cite{Sachdev, Dutta}
\begin{align}\label{eq:TFIM}
H_{\rm TFIM}(g)=-\omega\sum_{j=1}^N \left(g \sigma_j^x +\sigma_j^z\sigma_{j+1}^z\right),
\end{align}
with $N$ even and, for convenience, we assume periodic boundary conditions, $\sigma^{x,y,z}_{N+1}\!=\! \sigma^{x,y,z}_{1}$ where $\sigma^{x,y,z}$ are the Pauli matrices. Upon the standard Jordan-Wigner and Fourier transformations, Eq.~\eqref{eq:TFIM} decouples into a collection of independent Landau-Zener problems in momentum space~\cite{Dziarmaga2005,Sachdev,Dutta}, $H_{\rm TFIM}(g)\!\!=\!\!\bigoplus_{k>0} \Psi^\dagger_k H_{k}(g) \Psi_k$ with $\Psi^\dagger_k\!\!=\!\!(c_k^\dagger,c_{-k})$ the mode of Fourier-transformed fermionic operators and $H_{k}(g)=h_k^z(g)\sigma^k_z+h_k^x\sigma_x^k$, where $\sigma_{x,y,z}^k$ are the Pauli matrices for momentum $k$, which takes discrete values $k_n\!\!=\!\!(2n-1)\pi/(N b)$ with $n\!\!=\!\!1,\ldots,N/2$ and $b$ the distance between neighbouring spins. From the transformation, one finds the coefficients $h_k^z(g)\!\!=\!\!2\omega (g-\cos (k b))$, $h_k^x\!\!=\!\!2\omega \sin (k b)$, and the ground state energy for the $k$-momentum subspace reads as (see~\cite{SM} for details)
%
%
\begin{equation}
\label{eq:IsingEnergy}
\varepsilon_k(g)\!=\!-2\omega\sqrt{g^2+1-2g\cos(kb)}-2\omega g.
\end{equation}

There is a QPT at $g_c\!\!=\!\!1$~\cite{Sachdev} and for ramps that approach or cross the critical point in a finite time, $\tau_q$, we recover precisely the universal scaling laws from the predictions of the KZM~\cite{Zurek2005,Dziarmaga2005,Francuz2016}. Since the solution involves re-writing Eq.~\eqref{eq:TFIM} into independent Landau-Zener models in momentum space, determining the associated counterdiabatic Hamiltonian, Eq.~\eqref{eq:TQD}, becomes greatly simplified as it reduces to the concatenation of two-level controls~\cite{delCampoPRL2012, DamskiJStat, Berry2009, CampbellPRL2017}. We introduce the counterdiabatic driving, which for the $k$-subspace is $H_{k,{\rm CD}}(g(t))\!\!=\!\!\Theta_k(g(t))\sigma^k_y$, with $\Theta_k(g)\!\!=\!\!h_k^z(g)h_k^z(2 (h_k^{z,2}(g)+h_k^{x,2}))^{-1}$. We can readily determine Eq.~\eqref{eq:COST} for a given subspace, $\partial_t C\!\!=\!\!2|\Theta_k(g(t))|$ and the Bures angle between ground states at $g_0$ and $g(t)$, $\mathcal{L}_t\!\!=\!\!{\rm arccos}|\langle \psi_{\rm gs}(g(t)) | \psi_{\rm gs}(g_0)\rangle|\!\!=\!\!{\rm arccos}|\prod_{k>0} \cos(\theta_k(g(t))-\theta_k(g_0))|$, with  $\theta_k(g)\!\!=\!\!\arctan((h_k^z(g)-(h_k^{z,2}(g)+h_{k}^{x,2})^{1/2})/h_k^x)$~\cite{SM}.

We now have all the ingredients necessary to evaluate the QSL, Eq.~\eqref{eq:speednorm}, and to begin we consider a linear ramp, $g(t)\!\!=\!\!g_f t/\tau_q$. For low $k$ subspaces where $\varepsilon_k(g)\!\!\rightarrow\!\! 0$ for $g\!\!\rightarrow\!\! g_c$, and are therefore critical, the solid lines in Fig.~\ref{figTFIM}(a) show that the speed exhibits a behavior reminiscent of the adiabatic-impulse approximation. Indeed, notice that all lines fall on top of each other far from the critical point. Thus, the quantum speed is independent of the ramp duration, indicating the model is in the adiabatic regime. 

As the system approaches the critical point, and therefore crosses over into the impulse regime, we see quantitative differences emerge as we vary the ramp duration. Smaller $\tau_q$ leads to increased speeds in the vicinity of the QPT, while larger $\tau_q$ reduces the effective size of the critical region. This picture is consistent with the trade-off between the speed and the energetic cost of implementing quantum control~\cite{CampbellPRL2017, FunoPRL2017, XuPRA2018, Santos2015SciRep, Oleg2017PRL} and demonstrates that, near the critical point, control protocols are essential for achieving finite time adiabatic dynamics, while if the system is manipulated outside the impulse regime, where energy gaps are typically much larger, there is no need for complex control techniques. 

\begin{figure}[t]
\centering
\includegraphics[width=1.\linewidth,angle=0]{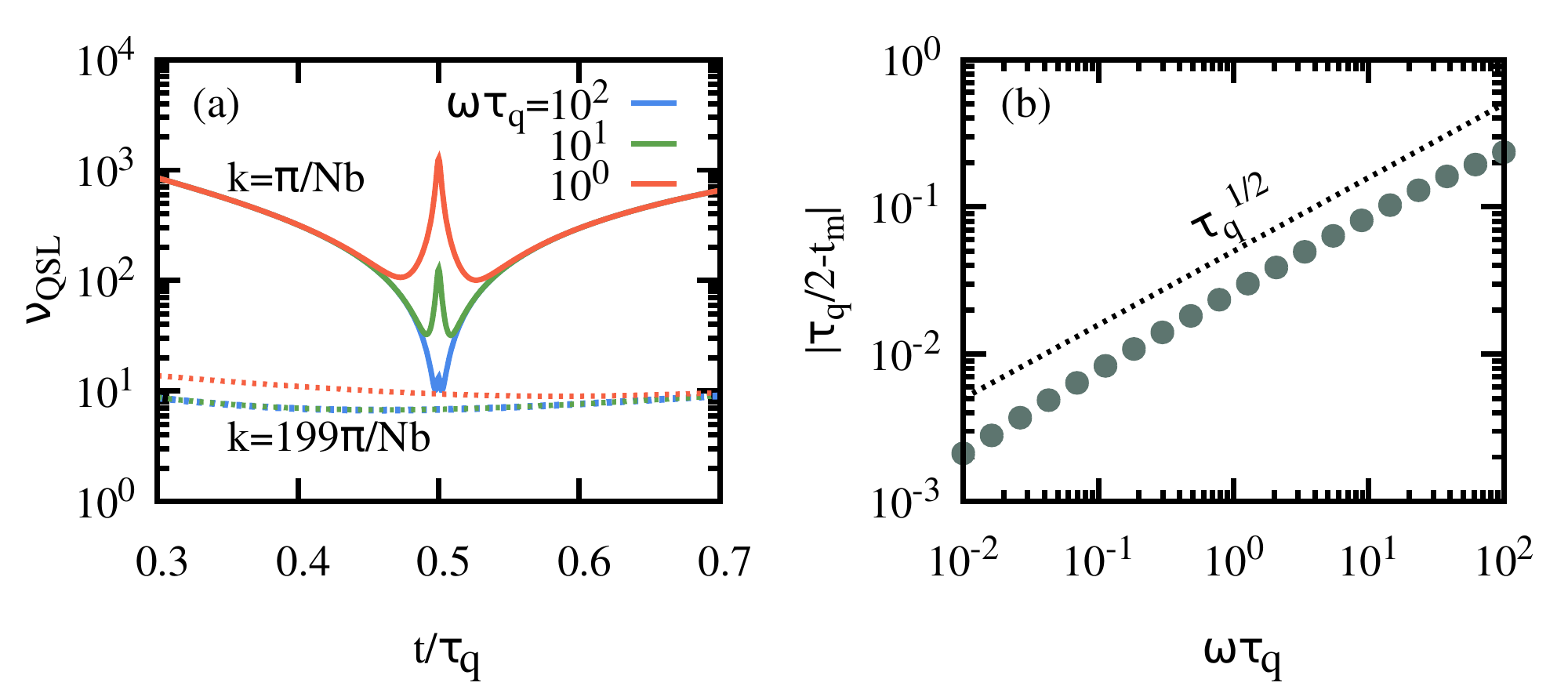}
\caption{\small{(a) Speed $\nu_{\rm QSL}$ as given in Eq.~\eqref{eq:speednorm} for the TFIM with counterdiabatic driving, quenched from $g(0)\!=\!0$ to $g(\tau_q)\!=\!2$ thus traversing the QPT and for different total ramp durations $\tau_q$ ($\omega\tau_q\!=\!10^2$, $10^1$ and $10^0$) and for two momenta $k=\pi/Nb$ (solid lines) and $k=199\pi/N b$ (dashed lines), with $N=1000$ spins. (b) Scaling of the duration of the impulse regime, i.e. of the time $|t_c-t_m|$, as a function of $\tau_q$, which shows KZM scaling $\tau_q^{z\nu/(1+z\nu)}\!=\!\tau_q^{1/2}$ (dashed line) for the TFIM as $z\nu\!=\!1$. Since $g(\tau_q)\!=\!2$, it follows $t_c\!=\!\tau_q/2$. See main text for further details.}}
\label{figTFIM}
\end{figure}

With the qualitative picture established, we now inspect the scaling properties of the instant of the time, $t_m$, at which $\nu_{\rm QSL}$ features a minimum. The intuition is clear: When no counterdiabitic control is necessary, the speed will be entirely dependent on the behavior of the ground state energy, cf. Eq.~\eqref{eq:speednorm}. To adiabatically cross the QPT without any additional control the speed must vanish at the critical point due to the closing energy gap. For finite time ramps, as the system approaches the QPT defects will become increasingly more likely to be generated and therefore the need for counterdiabatic control grows. For such a process, the transition from ``effectively adiabatic" to ``requires control" is reflected in the non-zero cost which in turn leads to increased speeds near the QPT~\cite{CampbellPRL2017}. Therefore the crossover from the adiabatic to impulse regimes is delineated by the minimum of $\nu_{\rm QSL}$, which takes place at a time instant $t_m$. We find that the duration of the impulse regime scales as $|t_c-t_m|\!\sim\! \tau^{1/2}$ with $t_c$ such that $g(t_c)\!\equiv \! g_c$,  which is in perfect agreement with the expected scaling $\hat{\tau}\!\!\sim\!\! \tau^{z\nu/(1+z\nu)}\!\!=\!\!\tau_q^{1/2}$ given in Eq.~\eqref{eq:AI} since $z\nu\!\!=\!\!1$ for the TFIM~\cite{Zurek2005,Dziarmaga2005}. This is shown in Fig.~\ref{figTFIM}(b) where fitting the points to a power law $\tau_q^\beta$ yields $\beta=0.51(1)$. Nonlinear protocols modify the KZM scaling in a non-trivial fashion. Nevertheless we have confirmed that for $g(t)\!\!=\!\!1-(t-t/\tau_q)^r$ and $g(t)\propto (t/\tau_q)^r$ the speed of the controlled dynamics still exhibits excellent agreement with $\hat{\tau}\!\!\sim\!\! \tau_q^{z\nu r/(1+z\nu r)}$ as one expects from KZM arguments~\cite{SM}.

If we turn our attention to higher momentum subspaces we find the critical scaling is lost. In Fig.~\ref{figTFIM}(a) the dashed lines correspond to a high momentum subspace with $k\!\!=199\pi/Nb$.  Such high momentum subspaces do not show any trace of the QPT, and we find a trivial scaling with ramp duration $\propto \tau_q$~\cite{SM}. This is indeed expected since these subspaces are not critical, and hence KZM arguments do not apply. The energy gap remains large throughout the evolution, and as a result, these subspaces do not contribute to the finite-time excitations in the bare nonequilibrium dynamics, making unnecessary the application of counterdiabatic control. This naturally emerges in our formalism: the trivial scaling in $\nu_{\rm QSL}$ in Eq.~\eqref{eq:speednorm} accounts for the absence of criticality, i.e., for the absence of competing energy scales between the resources to achieve controlled dynamics and the energetic change in the considered subspace. 

Finally, it is worth noting that KZM scaling was shown to be exhibited in the Landau-Zener problem in a seminal work by Damski~\cite{DamskiPRL2005} and therefore one may ask if we recover the same behavior in our setting. Due to its simplicity we can fully analytically treat the Landau-Zener problem and we find $t_m\!\!\sim\! \tau_q^{2/3}$ instead of the distinctive $\tau_q^{1/2}$ scaling for the TFIM~\cite{SM}. In contrast to Ref.~\cite{DamskiPRL2005}, no heuristic arguments are invoked here to determine the crossover between adiabatic and impulse regimes. Moreover, such a $\tau_q^{2/3}$ scaling is further supported from a numerical estimation of the crossover~\cite{SM}. Since both the TFIM and the LZ exhibit the same KZM critical exponents one naturally asks where the apparent discrepancy emerges. We find that the energy shift introduced when diagonalizing Eq.~\eqref{eq:TFIM}, i.e. the final term in Eq.~\eqref{eq:IsingEnergy}, is crucial to recovering the predicted KZM scaling exponent in the true many-body case.

\paragraph{Lipkin-Meshkov-Glick model.} As a final example we move to a more complex many body system. Originally introduced in the context of nuclear physics~\cite{Lipkin1965}, the LMG model has become the paradigmatic system to study extreme long-range interactions and their role in critical phenomena, both theoretically~\cite{Ribeiro:08,Ribeiro:07,Dusuel:04,Dusuel:05,Vidal:07,Zunkovic:18,FogartyPRL2020} and experimentally~\cite{Zibold:10,Zhang:17,Jurcevic:17}.
The Hamiltonian can be written as
\begin{align}\label{eq:LMG}
  H_{\rm LMG}(g)\! \! =\! -\omega J_z-\frac{g^2\omega}{N}J_x^2,
\end{align}
where $\omega$ sets the energy scale, while $g$ accounts for the relative strength of the ferromagnetic spin coupling, and $J_{\alpha}\!\!=\!\!\sum_{i=1}^N \sigma_i^\alpha/2$ for $\alpha\in \{x,y,z\}$. In the thermodynamic limit, $N\rightarrow\infty$, the LMG can be diagonalized via a Holstein-Primakoff transformation~\cite{SM}, which reveals a QPT at $g_c\!\!=\!\!1$~\cite{Ribeiro:07,Ribeiro:08,Dusuel:04,Dusuel:05}, 
\begin{equation}
\label{eq:LMGeff}
H_{\rm LMG,eff}(g)\!\!=\!\omega a^{\dagger}a-\frac{g^2\omega}{4}(a+a^\dagger)^2, 
\end{equation}
where $[a,a^\dagger]=1$ denote the bosonic excitations. The previous effective model is valid for $0\leq g\leq 1$.  Through an additional Bogoliubov transformation, Eq.~\eqref{eq:LMGeff} can be recast as a harmonic oscillator and therefore the corresponding counterdiabatic Hamiltonian is exactly known~\cite{Muga2010JPB, CampbellPRL2015}. Note that $H_{\rm LMG,eff}(g)$, also corresponds to a low-energy effective description of other critical models~\cite{Mottl:12,Hwang:15,Puebla:17,Peng:19,Puebla:19}. 

We consider again a linear ramp through the QPT according to $g(t)\!=\!t/\tau_q$ for $t\in[0,\tau_q]$. The speed of the controlled evolution, as defined in Eq.~\eqref{eq:speednorm}, is given by~\cite{CampbellPRL2017, SM}
\begin{align}\label{eq:nucostLMG}
\nu_{\rm QSL}=\frac{\sqrt{ \omega_t^2/4+(\dot{\omega}_t/(\sqrt{8}\omega_t))^2}}{\cos\mathcal{L}_t\sin\mathcal{L}_t}
\end{align}
with
\begin{equation}
\begin{split}
\mathcal{L}_t &={\rm arccos}\left(\sqrt{2\sqrt{\omega \omega_t}/(\omega +\omega_t)}\right)\\
&\text{and}\quad\omega_t=\omega\sqrt{1-g^2(t)}.
\end{split}
\end{equation}
In Fig.~\ref{figLMG}(a) we show the speed for different quench rates $\tau_q$. Similar to the TFIM, the rapid increase of $\nu_{\rm QSL}$ close to the QPT at $g\!\!=\!\!1$ suggests the presence of the two distinctive dynamical regimes predicted by the KZM. Computing the time, $t_m$, at which the speed features a minimum [cf. the solid points in Fig.~\ref{figLMG}(a)] we can analyze its scaling with the total quench time, $\tau_q$. From KZM arguments and for a linear ramp, one expects the uncontrolled dynamics, i.e. without counterdiabatic driving, to feature an impulse regime during a time $\hat{\tau}\!\sim\! \tau_q^{z\nu/(1+z\nu)}\!=\!\tau_q^{1/3}$ as $z\nu=1/2$ for the LMG universality class~\cite{Dusuel:04,Ribeiro:07,Hwang:15,Peng:19}. In Fig.~\ref{figLMG}(b) we confirm that the speed of the controlled dynamics exhibits the scaling $|\tau_q-t_m|\!\sim\!\tau_q^{1/3}$, which accounts for the duration of the impulse regime, precisely inline with the theoretical predictions. Indeed, a numerical fit to a power law $\tau_q^\beta$ in the interval $\omega\tau_q\in[10^3,10^5]$ leads to $\beta=0.34(1)$.
\begin{figure}[t]
\centering
\includegraphics[width=1.\linewidth,angle=0]{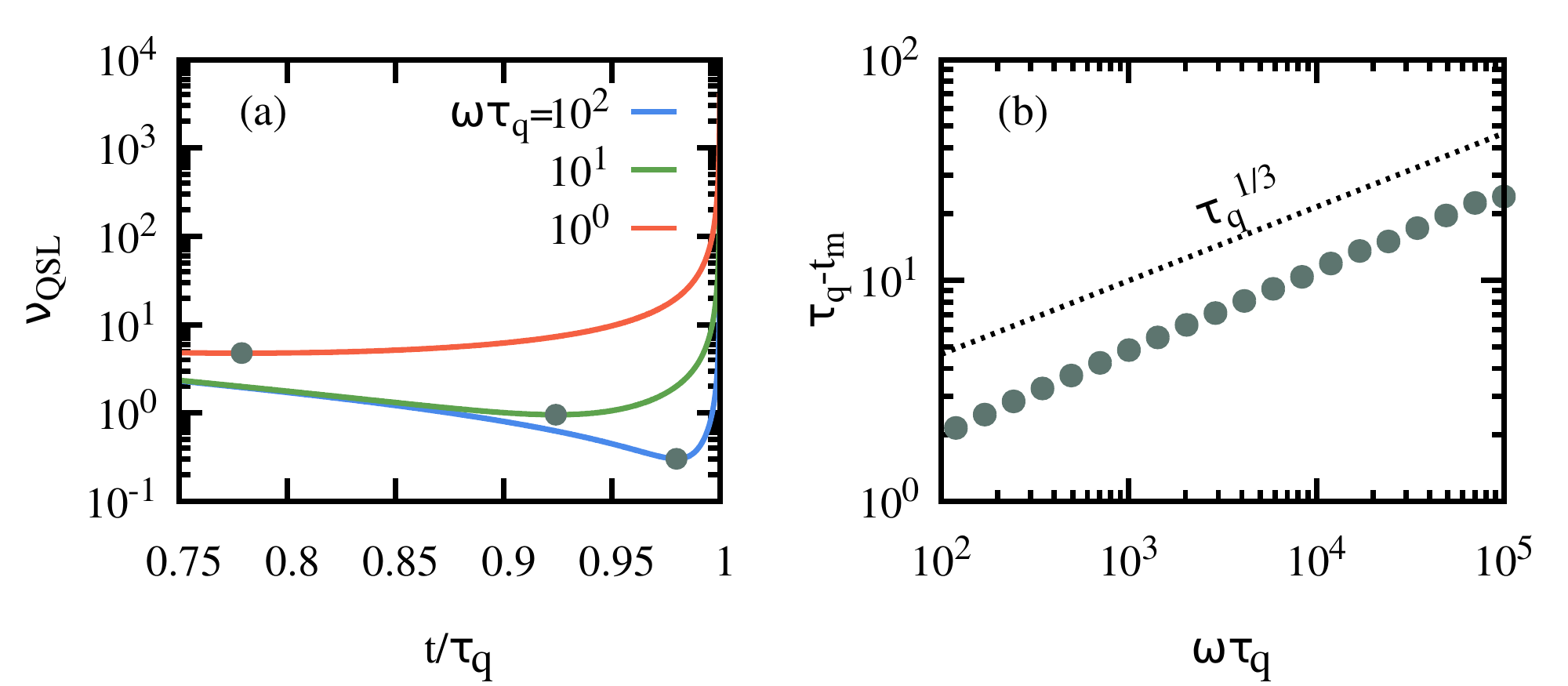}
\caption{\small{(a) Speed $\nu_{\rm QSL}$ as given in Eq.~\eqref{eq:speednorm} for the LMG model with counterdiabatic driving, quenched from $g(0)\!\!=\!\!0$ towards the QPT,  $g(\tau_q)\!\!=\!\!g_c\!\!=\!\!1$, for different total times $\tau_q$ ($\omega\tau_q\!\!=\!\!10^2$, $10^1$ and $10^0$). The minimum value of the speed, as well as its time location $t_m/\tau_q$, is indicated with solid points. (b) Scaling of the time $|t_c-t_m|$ with $t_c\!\!=\!\!\tau_q$ as $g(\tau_q)\!\!=\!\!g_c$, as a function of $\tau_q$, which shows the expected KZM scaling $\tau_q^{z\nu/(1+z\nu)}\!=\!\tau_q^{1/3}$ (dashed line) for the LMG as $z\nu=1/2$. See main text for details.}}
\label{figLMG}
\end{figure}

\paragraph{Concluding remarks.}
Quantum  error mitigation strategies are typically accompanied by a large computational or energetic overhead. The KZM provides a phenomenological framework to determine during which periods nonadiabatic excitations arise. Thus, in practice error mitigation needs only to be applied during these periods. In the present work, we have shown that this phenomenological prediction is encoded in the geometric QSL for counterdiabatic driving. Our findings crucially depend on the competition of the time-averaged energy of the bare Hamiltonian and the cost for counterdiabatic control. It is precisely this trade-off that contains the signatures of the nonequilibrium behavior. Thus, it is not far-fetched to realize that the paradigm of counterdiabatic driving may yet prove useful for adiabatic quantum computing. While complete control along the entire dynamics is generally unfeasible, the KZM demonstrates that only when the system is in the impulse regime are control fields needed. This period, that is the origin of all fundamentally non-correctable errors, is indicated by the critical behavior of the QSL.

\begin{acknowledgments}
We acknowledge insightful discussions with Adolfo del Campo. R. P. acknowledges the support by the SFI-DfE Investigator Programme (grant 15/IA/2864). This research was supported by grant number FQXi-RFP-1808 from the Foundational Questions Institute and Fetzer Franklin Fund, a donor advised fund of Silicon Valley Community Foundation (S. D.). S. C. gratefully acknowledges the Science Foundation Ireland Starting Investigator Research Grant ``SpeedDemon" (No. 18/SIRG/5508) for financial support.
\end{acknowledgments}

\bibliography{paper.bib}

%
%
%
%
%
%
%

\setcounter{equation}{0}
\setcounter{figure}{0}
\setcounter{table}{0}
\makeatletter
\renewcommand{\theequation}{S\arabic{equation}}
\renewcommand{\thefigure}{S\arabic{figure}}
\renewcommand{\bibnumfmt}[1]{[S#1]}

\begin{widetext}
\section{Supplemental Material \\ Kibble-Zurek scaling in quantum speed limits for shortcuts to adiabaticity}

\author{Ricardo Puebla}
\email{r.puebla@qub.ac.uk}
\affiliation{Centre for Theoretical Atomic, Molecular and Optical Physics, School of Mathematics and Physics, Queen’s University Belfast, Belfast BT7 1NN, United Kingdom}
\author{Sebastian Deffner}
\email{deffner@umbc.edu}
\affiliation{Department of Physics, University of Maryland, Baltimore County, Baltimore, MD, 21250, USA}
\author{Steve Campbell}
\email{steve.campbell@ucd.ie}
\affiliation{School of Physics, University College Dublin, Belfield Dublin 4, Ireland}

\maketitle

%
%

\section{Kibble-Zurek scaling from quantum speed limit}
As mentioned in the main text, the speed is given by
\begin{align}\label{eq:SMspeed}
    \nu_{\rm QSL}=\frac{\sqrt{\varepsilon_n^2(t)+(\partial_t C)^2}}{\cos(\mathcal{L}_t)\sin(\mathcal{L}_t)}.
\end{align}
Then, for critical systems the energy follows $\varepsilon(g)\sim |g-g_c|^{z\nu}$, while the cost of the counterdiabatic driving is given by $\partial_t C\sim |\dot{\varepsilon}/\varepsilon|$. By definition, a driven critical system when including a counterdiabatic term remains in the ground state at all times. The Bures angle is therefore only a function of $g$, that is, the denominator $D=\cos(\mathcal{L}_t)\sin(\mathcal{L}_t)$ in Eq.~\eqref{eq:SMspeed}  is independent of the time $\tau_q$ employed to perform the quench.

Assuming a linear quench $g(t)=t/\tau_q$ and taking $g_c=0$ by convenience, $\nu_{\rm QSL}$ takes the form 
\begin{align}
    \nu_{\rm QSL}(t)\sim \frac{\sqrt{(t/\tau_q)^{2z\nu}+(z\nu/t)^2}}{D}.
\end{align}
Note that the specific prefactors are not important to study the scaling properties, and have been neglected.  In particular, the previous expression features a minimum at a time $t_m$, which is given by 
\begin{align}
    t_m=(z\nu)^{1/(2(z\nu+1)}\tau_q^{z\nu/(1+z\nu)}\sim \tau_q^{z\nu/(1+z\nu)}
\end{align}
which directly follows from $d\nu_{\rm QSL}/dt=0$. Note that since $g(0)=g_c=0$, the previous time $t_m$ directly accounts for the duration of the impulse regime, which is precisely the Kibble-Zurek scaling prediction. Moreover, it is worth stressing that since $z\nu>0$ it follows $d^2\nu_{\rm QSL}/dt^2|_{t_m}>0$ for all $\tau_q$, ensuring the presence of a minimum. By inspecting the value of the speed at $t_m$ one immediately finds 
\begin{align}
    \nu_{\rm QSL,min}\equiv \nu_{\rm QSL}(t_m)\sim \tau_q^{-z\nu/(1+z\nu)}.
\end{align}

\section{Landau-Zener: more details}
Let us consider a Landau-Zener (LZ) model with a time-dependent magnetic field $g(t)$, whose Hamiltonian reads as
\begin{align}\label{eq:lz}
H_{\rm LZ}(t)=\Delta\sigma_x+g(t)\sigma_z.
\end{align}
In the following we provide the numerical estimation of the adiabatic-impulse scaling as well as the analytical derivation based on the approach explained in the main text.

\subsection{Numerical estimation of adiabatic-impulse scaling}
Here we provide an estimation of the width of the adiabatic-impulse regime from the numerical solution to the quenched dynamics. For that we compute the state at time $t$, $\ket{\psi(t)}$ following the evolution dictated by Eq.~\eqref{eq:lz} with $g(t)=g0+(g_1-g_0)t/\tau_q$ and $t\in[0,\tau_q]$ and initial state $\ket{\psi(t=0)}=\ket{\phi_0(g(0))}$, where $\ket{\phi_0(g)}$ denotes the ground state of $H_{\rm LZ}$ at $g$. In order to estimate the width of the impulse regime and the scaling thereof, we choose $g_0=-5\omega$ and $g_1=0$ while $\Delta=0.1\omega$, and compute the instantaneous infidelity, 
\begin{align}
    I(t)=1-|\langle \psi(t)| \phi_0(g(t)) \rangle |^2
\end{align}
so that $I(0)=0$. In the limit of $\omega\tau_q\rightarrow \infty$, the evolution is fully adiabatic and thus $I(t)=0 \ \forall t$. The location of the transition between adiabatic and impulse regime can be estimated as the instant $\hat{t}$ at which the instantaneous infidelity surpasses a certain threshold $K$, i.e. $I(t>\hat{t})\geq K$. Then, the duration of the impulse regime is given by $t^*=\tau_q-\hat{t}$. This is plotted in Fig.~\ref{figSM:LZ}(a), where $I(t)$ as a function of $t/\tau_q$ for different quench times $\tau_q$. In Fig.~\ref{figSM:LZ}(b) we show the values $t^*$ obtained for three different choices of $K$, namely, $K=10^{-2}$, $10^{-3}$ and $10^{-4}$, as a function of $\omega\tau_q$. 
A fit to a power-law $a\tau_q^b$ leads to $b=0.669(4)$ for $K=10^{-5}$ and $\omega\tau_q\in [10^1,10^3]$. For the other two chosen thresholds, $K=10^{-3}$ and $K=10^{-4}$, we get $b=0.65(2)$ and $b=0.674(8)$ , respectively. These results strongly suggest that the duration of the impulse regime, $t^*$, scales as $\tau_q^{2/3}$. This is further corroborated with the analytical derivation provided in the next part. 

\begin{figure}
\centering
\includegraphics[width=0.75\linewidth,angle=0]{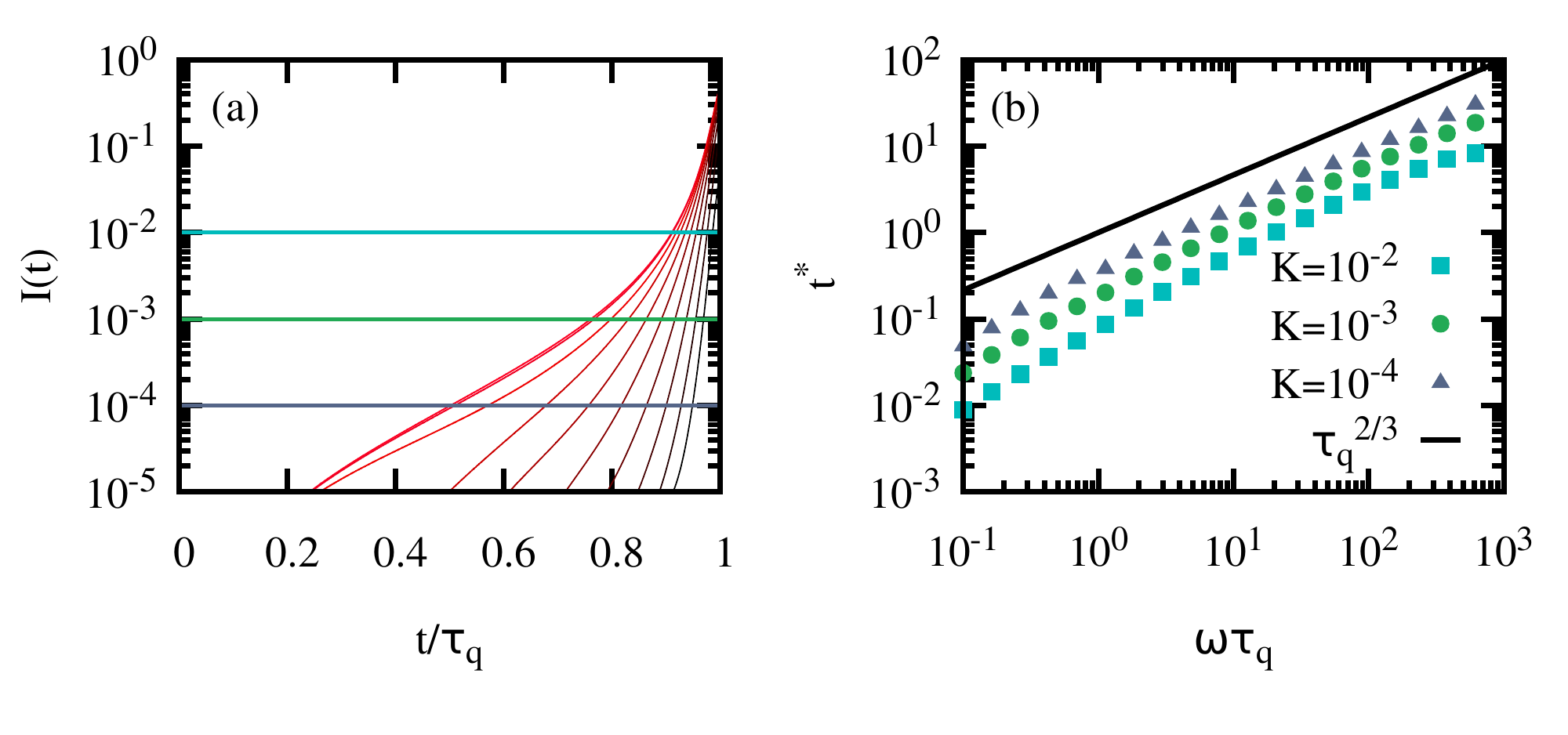}
\caption{\small{(a) Instantaneous infidelity $I(t)$ for the LZ with $\Delta=0.1\omega$, $g_0=-5\omega$ and $g_1=0$, and for different quench times, from $\omega\tau_q=10^{-1}$ (red line) to $10^3$ (black line). Horizontal lines illustrate the threshold $K$ to estimate the transition between adiabatic and impulse regimes. The time at which the transition takes place, $\hat{t}$, is estimated as  $I(\hat{t}>t)>K$.  (b) Scaling of $t^*=\tau_q-\hat{t}$ as a function of the quench time $\tau_q$ obtained for different $K$. The results indicate the scaling $t^*\sim \tau_q^{2/3}$.}}
\label{figSM:LZ}
\end{figure}

\subsection{Analytical derivation of the adiabatic-impulse scaling}
Now, for convenience we consider that $g(t)= g_1 t/\tau_q$ running from $t\in[0,\tau_q]$, so that the initial state is the ground state of $H_{\rm LZ}$ with $g(0)=0$.  The energy gap of the LZ model is given by
\begin{align}
\epsilon_{\rm LZ}(g)=\varepsilon_{1}(g)-\varepsilon_0(g)=2 \left(\Delta^2+g^2 \right)^{1/2}.
  \end{align}
The counterdiabatic Hamiltonian for the LZ is given by
\begin{align}
H_{\rm CD}(t)=\theta(t)\sigma_y, \qquad \qquad {\rm with}\qquad \theta(t)=-\frac{\dot{g}(t)\Delta}{2(\Delta^2+g^2(t))}=-\frac{2\delta_g\Delta }{\tau_q\epsilon_{\rm LZ}^2(g(t))}.
  \end{align}
The instantaneous cost $\partial_t C=|| H_{\rm CD}(t)||$ then is simply given by $\partial_t C=2|\theta(t)|$. For $g(t)= g_1 t/\tau_q$, the expression for the speed simplifies to
\begin{align}
\nu_{\rm QSL}=2\left[\frac{g_1^2t^2}{\tau_q^2}+\Delta^2\left(1 +\frac{g_1^2\tau_q^2}{(g_1^2t^2+\Delta^2\tau_q^2)^2}\right) \right]^{1/2}\sin^{-1}\left(\frac{\Delta-g_1+\sqrt{\Delta^2+g_1^2}}{\sqrt{\Delta^2+g_1(g_1-\sqrt{\Delta^2+g_1^2})}}\right)\nonumber
  \end{align}
From the previous expression we find
\begin{align}
  t_{m}=\left(\frac{2^{1/3}\Delta^{2/3}\tau_q^{4/3}}{g_1^{4/3}}-\frac{\Delta^2\tau_q^2}{g_1^2} \right)^{1/2}\approx \frac{2^{1/6}\Delta^{1/3}\tau_q^{2/3}}{g_1^{2/3}}+O(\Delta^{5/3})\nonumber
  \end{align}
provided $\tau_q \leq \tau_q^*= \sqrt{2}g_1 \Delta^{-2}$, where $\tau_q^*$ sets the maximum quench time such that $\nu_{\rm QSL}$ features a minimum at the given time $t_{m}$. From the previous derivation it follows that $t_{m}\sim \tau_q^{2/3}$ as suggested by the numerical analysis shown above.

\section{One dimensional transverse-field Ising model: more details}

Let us consider now the paradigmatic  one-dimensional Ising model with transverse field, whose Hamiltonian can be written as
\begin{align}
H_{\rm TFIM}=-\omega\sum_{j=1}^N \left(g \sigma_j^x +\sigma_j^z\sigma_{j+1}^z\right),
\end{align}
with $N$ even and periodic boundary conditions for convenience, $\sigma^{x,y,z}_{N+1}=\sigma^{x,y,z}_{1}$. This model features a QPT at $g_c=1$~\cite{Sachdev}. Upon a Jordan-Wigner transformation and a Fourier transformation, the Hamiltonian $H_{\rm TFIM}$ decouples in the momentum space as a set of $N$ Landau-Zener problems $H_k$ with different energy splitting among the $k$ quasiparticles (or $N/2$ if split according to the $\mathbb{Z}_2$ parity symmetry), $H_{\rm TFIM}=\bigoplus_{k>0} \Psi^\dagger_k H_{k} \Psi_k$ with $\Psi^\dagger_k=(c_k^\dagger,c_{-k})$ the mode of Fourier-transformed fermionic operators~\cite{Dziarmaga2005,Sachdev,Dutta}, so that
\begin{align}
    H_k=h_k^z(g) \sigma_z^k+h_k^x \sigma_x^k,
\end{align}
and $\sigma_z^k=\ket{1}_k\bra{1}_k-\ket{0}_k\bra{0}_k$, 
$h_k^z(g)=2\omega (g-\cos (k b))$, $h_k^x=2\omega\sin (k b)$, and allowed wavenumber $k_n=(2n-1)\pi/(N b)$ with $n=1,\ldots,N/2$, where $b$ is the spacing between spins. One can diagonalize $H_k$, such that
\begin{align}\label{eq:HkIsing}
H_{k}=\epsilon_k(g)\ket{\phi_{k,+}(g)}\bra{ \phi_{k,+}(g)}-\epsilon_k(g)\ket{\phi_{k,-}(g)}\bra{\phi_{k,-}(g)}
\end{align}
with 
\begin{align}
\epsilon_k(g)=2\omega\sqrt{g^2+1-2g\cos(kb)}
\end{align}
and 
\begin{align}
    \ket{\phi_{k,+}(g)}&=\sin \theta_k(g) \ket{0}_k-\cos\theta_k(g)\ket{1}_k\\
    \ket{\phi_{k,-}(g)}&=\cos \theta_k(g) \ket{0}_k+\sin\theta_k(g)\ket{1}_k
\end{align}
such that $\theta_k(g)=\arctan(h_k^z(g)/h_k^x-\sqrt{h_k^{z,2}(g)+h_{k}^{x,2}}/h_k^x)$. It is important to note that each $k$ subspace carries a constant energy shift of $-2\omega g$. 

Since the TFIM decouples in a set of LZ problems, we can readily apply the expressions we obtained before. The ground state of the Ising model can be written as a tensor product
\begin{align}
|\psi_{\rm gs}(g)\rangle=\bigotimes_{k>0} \ket{\phi_{k,-}(g)}=\bigotimes_{k>0}\left(\cos \theta_k(g) \ket{0}_k+\sin\theta_k(g)\ket{1}_k\right).
\end{align}
Recall that $\ket{0}_k$ and  $\ket{1}_k$ denote zero and two quasi-particles with momentum $k$ and $-k$, respectively, i.e. $\ket{1}_k\equiv \ket{k,-k}$. 

Since the speed limit is given by
\begin{align}
\nu_{\rm QSL}=\frac{\sqrt{\varepsilon^2(t)+(\partial_t C)^2}}{\cos\mathcal{L}_t\sin\mathcal{L}_t}
  \end{align}
  we need to compute the ground-state energy $\varepsilon(t)$, the instantaneous cost $\partial_t C$ and the Bures angle $\mathcal{L}_t$. In particular, the ground-state energy reads as
\begin{align}
  \varepsilon(t)=-\sum_{n=1}^{N/2}\epsilon_{k_n}(g(t))-2\omega g=-2
  \omega \sum_{n=1}^{N/2}\sqrt{g^2(t)+1-2g(t)\cos k_n b}-2\omega g
  \end{align}.
  Note that we have included the constant energy shift $-2\omega g$. The cost is given by
\begin{align}
  \partial_t C=\sum_{n=1}^{N/2} 2 | \Theta_{\rm CD,k_n}(t)|=\sum_{n=1}^{N/2}\frac{|\dot{h}_{k_n}^z(g(t))h_{k_n}^x|}{h_{k_n}^{z,2}(g(t))+h_{k_n}^{x,2}}=\sum_{n=1}^{N/2}\frac{|\dot{g}(t)\sin k_n b|}{|1-2g(t)\cos k_nb+g^2(t)|}
\end{align}
where $\Theta_{\rm CD,k}(t)$ is the amplitude of the CD Hamiltonian in the $k$-subspace, which takes the form $H_{\rm CD,k}=\Theta_{\rm CD,k}(t)\sigma_y^{k}$ (see above for the Landau-Zener).

The overlap between ground states is given by
\begin{align}
\langle \psi_{\rm gs}(g) | \psi_{\rm gs}(g')\rangle = \prod_{k>0} \cos(\theta_k(g)-\theta_k(g'))
  \end{align}
so that $\mathcal{L}_t={\rm arccos}|\langle \psi_{\rm gs}(g(t)) | \psi_{\rm gs}(g_0)\rangle|={\rm arccos}|\prod_{k>0} \cos(\theta_k(g(t))-\theta_k(g_0))|$.

\section{Lipkin-Meshkov-Glick model: More details}
The Lipkin-Meshkov-Glick (LMG) model~\cite{Lipkin1965,Ribeiro:08,Ribeiro:07,Dusuel:04} can be written as
\begin{align}
H_{\rm LMG}(g)=-\frac{\omega}{2}\sum_i \sigma_i^z-\frac{g^2\omega}{2N}\sum_{i\leq j}\sigma_i^x\sigma_j^x
  \end{align}
where the $\sigma_{i}^{x,y,z}$ represent the spin-$\frac{1}{2}$ Pauli matrices of each of the $N$ interacting spins, with an energy scale $\omega$. The dimensionless parameter $g$ accounts for the relative strength of the ferromagnetic spin coupling. Introducing the $N$-spin representation, i.e., $J_{x,y,z}$ such that $[J_i,J_j]=i\epsilon_{ijk}J_k$, with $J_{\alpha}=\sum_{i=1}^N\sigma_i^\alpha/2$ for $\alpha\in \{x,y,z\}$ the Hamiltonian becomes
\begin{align}
  H_{\rm LMG}(g)=-\omega J_z-\frac{g^2\omega}{N}J_x^2.
\end{align}
In thermodynamic limit, $N\rightarrow \infty$, this model shows a QPT at $g_c=1$~\cite{Ribeiro:08,Ribeiro:07,Dusuel:04,Dusuel:05}.
As $J^2$ commutes with $H_{\rm LMG}(g)$, we constrain ourselves to the subspace of maximum angular momentum $J_N=N/2$. Making use of the Holstein-Primakoff transformation, $J_z=J_N-a^{\dagger}a$, $J_+=2J_N\sqrt{1-a^{\dagger}a/(2J_N)}a$, and $J_-=2J_N a^{\dagger}\sqrt{1-a^{\dagger}a/(2J_N)}$, with $J_x=(J_++J_-)/2$, the Hamiltonian becomes
\begin{align}\label{SM_eq:HLMG}
  H_{\rm LMG,eff}(g)=\omega a^{\dagger}a-\frac{g^2\omega}{4}(a+a^\dagger)^2
\end{align}
upon taking the $N\rightarrow \infty$ limit, i.e. $J_{+}\approx 2J_Na$ and $J_{-}\approx 2J_N a^{\dagger}$. The previous Hamiltonian is valid for $0\leq g\leq 1$. For $g>1$, a rotation of $J_{x,y,z}$ is required to properly set the quantization axis in the Holstein-Primakoff transformation.  The Eq.~\eqref{SM_eq:HLMG} is nothing but an harmonic oscillator with a $x^2$ perturbation.

For $g(t)=t/\tau_q$ (i.e. quench towards the critical point), the speed as defined in~\cite{CampbellPRL2017}, can be applied here too:
\begin{align}\label{eq:nucostLMG}
\nu_{\rm QSL}=\frac{\sqrt{ \omega_t^2/4+(\dot{\omega}_t/(\sqrt{8}\omega_t))^2}}{\cos\mathcal{L}_t\sin\mathcal{L}_t}
  \end{align}
with $\mathcal{L}_t={\rm arccos}\left(\sqrt{2\sqrt{\omega \omega_t}/(\omega +\omega_t)}\right)$ and $\omega_t=\omega\sqrt{1-g^2(t)}$. 

\begin{figure}
\centering
\includegraphics[width=0.75\linewidth,angle=0]{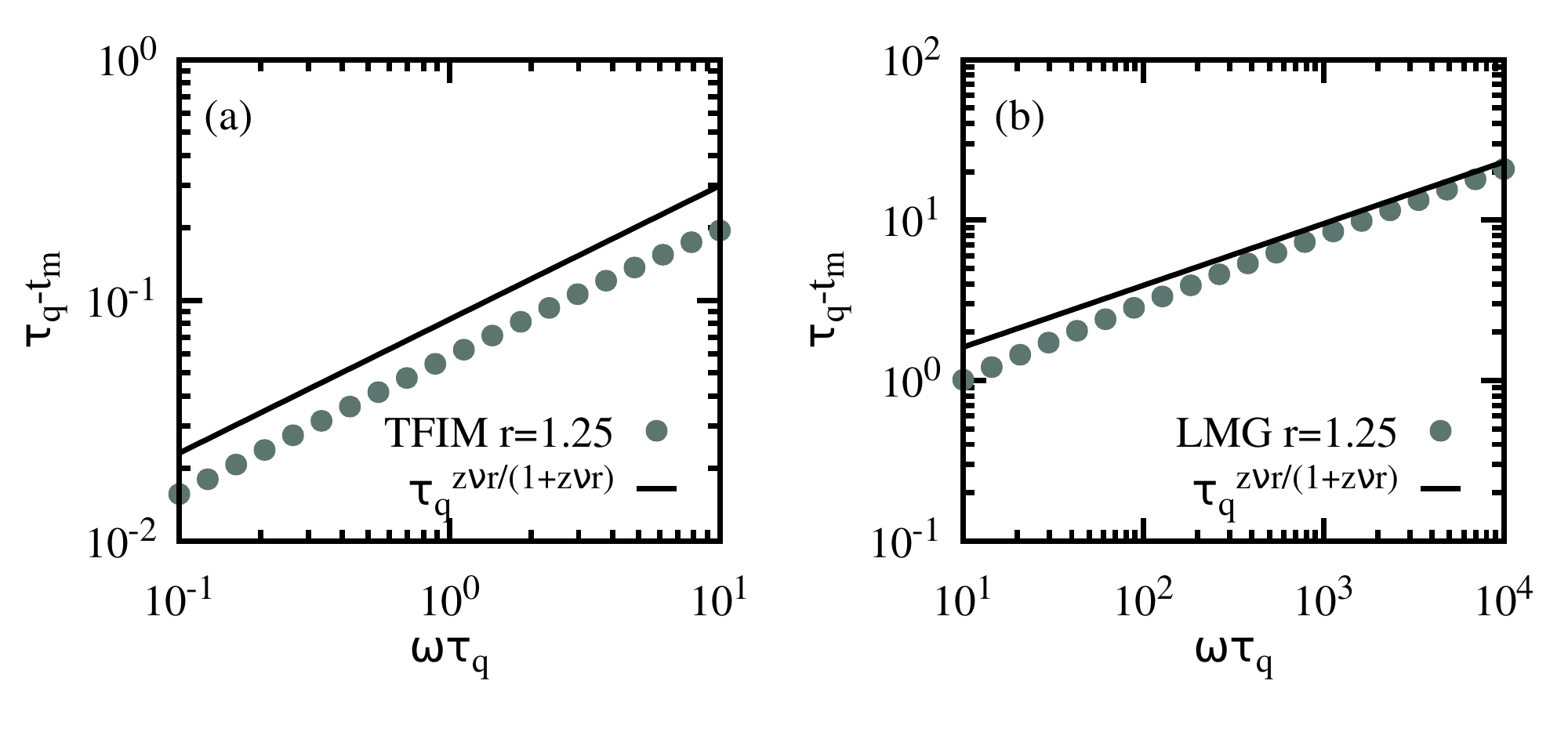}
\caption{\small{Scaling of the impulse regime based on the time $t_m$ at which $\nu_{\rm QSL}$ shows a minimum for nonlinear quenches, with $r=5/4$, and for the TFIM $z\nu=1$ (a) and LMG $z\nu=1/2$ (b). Lines are guides to the eyes displaying the expected Kibble-Zurek scaling. In (a) the points have been obtained for $N=1000$ spins and $k=\pi/N b$, while (b) is obtained in the thermodynamic limit.}}
\label{figSM:nonlin}
\end{figure}

\section{Nonlinear quenches}

As mentioned in the main text, for non-linear quenches approaching the QPT, $g(t)=1-(1-t/\tau_q)^r$, we obtain a good agreement with $|t_c-t_{m}|\sim \tau^{z\nu r/(1+z\nu r)}$ as one expects from KZM arguments. Recall that $t_c$ is the time at which $g(t_c)\equiv g_c$. Under the nonlinear protocol for $0\leq t\leq \tau_q$, it follows $t_c=\tau_q$. In order to verify this we consider quenches toward the QPT at different rates, and compute again the time $t_m$ at which the speed shows a  minimum.  In Fig.~\ref{figSM:nonlin} we show an example with $r=5/4$ for the TFIM and the LMG.

A fit of the points shown in Fig.~\ref{figSM:nonlin}(a) to a power law $\tau_q^\beta$ leads to $\beta=0.551(3)$ in the region $\omega\tau_q\in [10^{-1},10^1]$, which is in very good agreement with the expected $z\nu r/(1+z\nu r)=5/9$ for the TFIM. In a similar manner, for the LMG (cf. Fig.~\ref{figSM:nonlin}(b)) we obtain a value close to the expected $z\nu r/(1+z\nu r)=5/13$ when fitting the results in the interval $\omega\tau_q\in [10^3,10^4]$. Indeed, by performing the fit for slower quenches, $\omega\tau_q\in [10^4,10^6]$ we obtain a better agreement with the expected Kibble-Zurek scaling as $\beta=0.388(5)$. Similar results are obtained for different nonlinear exponents $r$.


\end{widetext}


\end{document}